\documentclass[runningheads]{llncs}
\usepackage{amsmath}
\usepackage{amssymb}
\usepackage{graphicx}
\usepackage{xcolor}
\usepackage{svg}
\usepackage{bm}

\usepackage{todonotes}
\newcommand{\repeatthanks}{\textsuperscript{\thefootnote}}

\begin{document}
\title{Global Control for Local SO(3)-Equivariant Scale-Invariant Vessel Segmentation}
\titlerunning{SO(3)-equivariant scale-invariant vessel segmentation}

\author{
Patryk~Rygiel\thanks{Equal contribution.}\inst{1}\orcidID{0009-0003-8539-5581} 
\and Dieuwertje~Alblas\repeatthanks\inst{1}\orcidID{0000-0002-7754-7405} 
\and Christoph~Brune\inst{1}\orcidID{0000-0003-0145-5069} 
\and Kak~Khee~Yeung\inst{2,3}\orcidID{0000-0002-8455-286X} 
\and Jelmer~M.~Wolterink\inst{1}\orcidID{0000-0001-5505-475X}
}
\authorrunning{P. Rygiel et al.}

\institute{
Department of Applied Mathematics, Technical Medical Centre, University of Twente, Enschede, The Netherlands \\ \email{\{p.t.rygiel,d.alblas,c.brune,j.m.wolterink\}@utwente.nl}
\and
Department of Surgery, Amsterdam UMC location Vrije Universiteit Amsterdam, Amsterdam, The Netherlands
\and
Amsterdam Cardiovascular Sciences, Microcirculation, Amsterdam, The Netherlands
}

\maketitle

\begin{abstract}
Personalized 3D vascular models can aid in a range of diagnostic, prognostic, and treatment-planning tasks relevant to cardiovascular disease management. Deep learning provides a means to obtain such models automatically from image data. Ideally, a user should have control over the included region in the vascular model. Additionally, the model should be watertight and highly accurate. To this end, we propose a combination of a global controller leveraging voxel mask segmentations to provide boundary conditions for vessels of interest to a local, iterative vessel segmentation model. We introduce the preservation of scale- and rotational symmetries in the local segmentation model, leading to generalisation to vessels of unseen sizes and orientations. Combined with the global controller, this enables flexible 3D vascular model building, without additional retraining. We demonstrate the potential of our method on a dataset containing abdominal aortic aneurysms (AAAs). Our method performs on par with a state-of-the-art segmentation model in the segmentation of AAAs, iliac arteries, and renal arteries, while providing a watertight, smooth surface representation. 
Moreover, we demonstrate that by adapting the global controller, we can easily extend vessel sections in the 3D model. Our code is available on GitHub\footnote{https://github.com/MIAGroupUT/SIRE-segmentation}.


\end{abstract}

\section{Introduction}

Detailed and topologically correct 3D vascular models can aid in many diagnostic, prognostic, and treatment-planning tasks in patients with cardiovascular diseases. Examples include diameter measurements in aneurysms~\cite{aaa-treatment}, tortuosity measurements~\cite{artery_tortuosity}, fitting stent grafts ~\cite{larrabide2008fast}, or hemodynamic analysis with computational fluid dynamics (CFD)~\cite{TAYLOR2023116414}. Automatically obtaining 3D models for these purposes has long been a topic of interest~\cite{lesage2009review} that has been primarily targeted with deep learning (DL) recently~\cite{chen2020deep,Litjens2019-he}.

Most current DL methods for 3D vessel segmentation perform voxel segmentation.
These models interpret the \textit{global} context of the image and yield a 3D voxel mask of the vessels, e.g.,~\cite{gharleghi2022automated,van2024automated}. 
However, the topological correctness of such voxel masks can only be imposed as a soft constraint \cite{cldice}. 
Moreover, the resolution of voxel masks is limited to that of the scanner, which for downstream tasks such as CFD is often insufficient. Additionally, expanding the segmentation's region of interest (ROI) requires adapting annotations and retraining the network, limiting the flexibility of voxel-based segmentation. Alternatively, vessels can be parametrized as generalized cylinders consisting of a manually or automatically~\cite{centerline-wolterink,friman2010multiple,sire-alblas} identified centerline and a set of \textit{locally} orthogonal contours \cite{shani1984splines}. This parametrization has been successfully integrated into machine learning approaches that achieve sub-voxel accuracy~\cite{carotids-alblas,lugauer2014improving,wolterink2019graph}. Performing segmentation locally allows for the incorporation of rotational and scale symmetries present in vessels~\cite{carotids-alblas,sire-alblas} to learn efficiently from few annotations. In contrast to voxel masks, a generalized cylinder is by design a topologically correct representation of the vessel and allows for expanding the segmentation's ROI by elongating the centerline. However, a drawback of local tracking and segmentation is the lack of global context, making it difficult to detect the start and end of the desired vascular centerline. When the user is interested in only a particular segment of the vasculature, stopping criteria based on local heuristics~\cite{friman2010multiple,van2022untangling,centerline-wolterink} are unable to constrain a local model without global information.

In this work, we unify global and local segmentation approaches by introducing a global controller for a local segmentation model. Open source multi-organ segmentation models like TotalSegmentator~\cite{total-segmentator} are a useful commodity for the community that we here leverage to provide a set of \textit{global} boundary conditions to constrain \textit{local} tracking and segmentation. This gives us the best of both worlds: we can select only those parts of the vasculature that are relevant to a particular task, and we can exploit local symmetries in vessel segmentation to obtain generalizable models with sub-voxel accuracy. Specifically, we introduce scale-invariant and rotation-equivariant local segmentation, which generalizes between vessels with very different calibres and tortuosities. We reconstruct smooth and watertight surfaces for each artery by fitting a neural field, that represents signed distance function (SDF), to the contours resulting from the local segmentation and blend these surfaces into watertight meshes of the vasculature \cite{alblas2022going}. We demonstrate the effectiveness of our method on a dataset of patients with abdominal aortic aneurysms (AAAs), where we obtain topologically correct and smooth models of the renal arteries, iliac arteries, and abdominal aorta. Moreover, we show how the user can efficiently adapt the segmentation ROI. Our results show that global and local segmentation are not mutually exclusive, but can be combined in a fruitful synergy to add value for cardiovascular disease management.

\begin{figure}[!t]
    \centering
    \includegraphics[width=\textwidth]{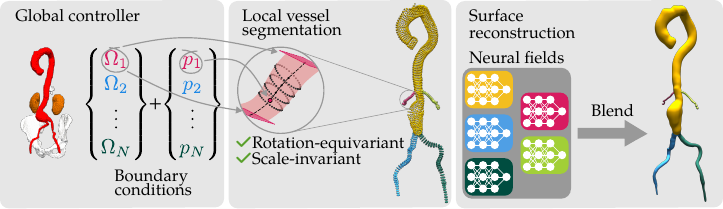}
    \caption{Proposed framework. 
    A global segmentation or localization method is used to provide boundary conditions $[\Omega_i, p_i]$,  such as points or regions-of-interest (ROIs) for the global control of local vessel segmentation. 
    A local joint tracking and segmentation model accurately segments vessels in a rotation-equivariant, scale-invariant manner. 
    Obtained contour sets are reconstructed into a watertight surface using neural fields and blended into a single vascular model \cite{alblas2022going}.}
    \label{fig:pipeline}
\end{figure}

\section{Materials \& Methods}
Figure~\ref{fig:pipeline} visualizes our framework, consisting of three elements: (1) extraction of boundary conditions for a global controller, (2) iterative local vessel segmentation subject to these boundary conditions, and (3) reconstruction of a smooth watertight surface i.e., a mesh without self-intersections, nor non-manifold edges or vertices, in which each edge has exactly two incident vertices.

\subsection{Global controller}
Iterative local vessel trackers lack a global overview and hence do not know where to start or terminate. 
In previous works, starting points were sampled from a segmentation mask, and stopping criteria were based on heuristics intrinsic to the tracker~\cite{sire-alblas,centerline-wolterink}, but this often resulted in early stopping or roaming of the tracker. 
Instead, we use coarse segmentations of TotalSegmentator \cite{total-segmentator}, an off-the-shelf multi-organ segmentation algorithm, as a global controller for the precise and topologically accurate iterative vessel segmentation module. For each vessel of interest $v_i$ in our 3D vascular model, we use this coarse segmentation to determine the boundary conditions, i.e. a seed point $p_i$ and the set of stopping criteria $\Omega_i$. Seed points are represented as 3D points and stopping criteria are represented as 3D binary masks. The boundaries of these masks define the regions in which segmentation can be performed; hence, crossing these boundaries causes the tracker to stop.

\subsection{Local vessel tracking and segmentation}
\begin{figure}[!t]
    \centering
    \includegraphics[width=\textwidth]{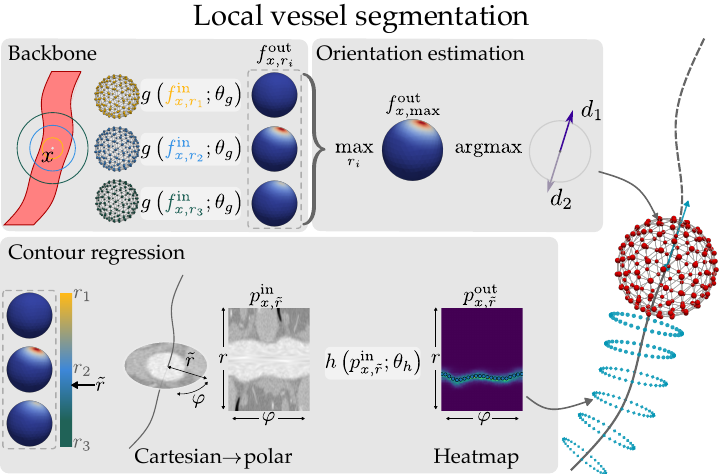}
    \caption{Overview of our iterative segmentation method, consisting of two steps. (1) At a point $x$, local vessel orientations are determined based on multi-scale spherical image features. (2) Based on the multi-scale responses, the optimal scale $\tilde{r}$ is selected, from which a scale-adaptive polar image normal to the local vessel orientation is constructed. CNN $h(\cdot; \theta_h)$ predicts lumen radii for each angle in $p^{\text{in}}_{\tilde{r}}$, forming a closed contour in Cartesian space.}
    \label{fig:method}
\end{figure}
From a local perspective, vessel appearance is subject to change due to varying orientations and diameters of the vessel. For example, the diameters of renal arteries are around 5 mm, iliac arteries range around 10 mm, while AAAs with diameters of $\>$80 mm are not uncommon \cite{aggarwal2011abdominal}. Moreover, renal arteries are generally more tortuous than the abdominal aorta. This poses a challenge to local artery segmentation methods that use image patches of fixed resolution, size, and orientation,e.g.~\cite{carotids-alblas,lugauer2014improving,wolterink2019graph}. A local segmentation method that is unaffected by vessel diameter and tortuosity is thus desirable.

We propose a scale-invariant, rotation-equivariant local segmentation module that performs iterative joint tracking and segmentation of the vessel, using its cylindrical shape as an anatomical prior. To track vessel $v_i$, the module is initialized at seed point $p_i$, provided by the global controller. First, the module estimates up- and downstream vessel directions $\bm{d}_1$ and $\bm{d}_2$ at $p_i$. Subsequently, the module takes a step of size $\Delta$ in direction $\bm{d}_1$. 
Every $\eta$ steps, the module delineates the lumen boundary orthogonal to the local vessel orientation. These two steps are shown in Figure \ref{fig:method}. Once one of the boundary conditions in $\Omega_i$ is violated, the module terminates. If direction $\bm{d}_2$ is not yet explored, the module is re-initialized at $p_i$ and traverses the vessel in direction $\bm{d}_2$ until termination. Both resulting centerline paths are combined into one connected centerline.


Both steps of the iterative segmentation rely on a scale-invariant, rotation-equivariant backbone \cite{sire-alblas}, which extracts features from image data centered around a point $x\in \mathbb{R}^3$ (top-left Fig. \ref{fig:method}).
This backbone consists of a graph convolutional network (GCN) ~\cite{gem-cnn-haan} $g(\cdot ; \theta_g)$ operating on the surface of the unit sphere on which local image information within a predefined radius $r$ of $x$ has been projected ($f^{\text{in}}_{x, r}$).
The backbone outputs a scalar field $f^{\text{out}}_{x, r}$ on the unit sphere.
To adapt to vessels of different calibres, the backbone processes local image information at different scales in parallel. We construct a set of multi-scale inputs $\{f^{\text{in}}_{x, r_j}\}_{j=1,...,m}$ from a pre-defined set of $m$ scales $R=\{r_j\}_{j=1,...,m}$ that we feed through $g(\cdot;\theta_g)$. This results in a set of multi-scale outputs $\{f^{\text{out}}_{x, r_j}\}_{j=1,...,m}$, which are used by the orientation estimation and contour regression modules.

The vessel orientation estimation module first aggregates $\{ f^{\text{out}}_{x, r_j}\}_{j=1,...,m}$ into a scale-invariant output $f^{\text{out}}_{x, \text{max}}$ by using a permutation-invariant maximum operation across the scales (top-right Fig. \ref{fig:method}) \cite{sire-alblas}. 
The local orientations $\bm{d}_1$ and $\bm{d}_2$ of the vessel at $x$ are represented by the locations of $f^{\text{out}}_{x, \text{max}}$, at least 90\textsuperscript{$\circ$} apart. The vessel orientation estimation module is trained with an MSE loss between $f^{\text{out}}_{x, \text{max}}$ and a target heatmap as in~\cite{sire-alblas}.

The contour regression module yields a closed contour delineating the lumen boundary on a 2D plane through $x$ orthogonal to the centerline, while adapting to the vessel calibre (bottom Fig. \ref{fig:method}). 
Closed contours are enforced by estimating the lumen radius for each angle on a polar transformed image $p^{\text{in}}_{x,\tilde{r}} \in \mathbb{R}^{N_{r}\times N_{\varphi}}$, centered at $x$ with radius $\tilde{r}$ \cite{carotids-alblas}. 
Instead of using polar images with fixed $\tilde{r}$ as in \cite{carotids-alblas}, we adapt the radius $\tilde{r}$ to the local vessel size, by computing it as the weighted average of the maximum activations on each scale-wise response in $\{f^{\text{out}}_{x, r_j} \}_{j=1,...,m}$. Based on $p^{\text{in}}_{x,\tilde{r}}$, a CNN $h(\cdot ; \theta_h)$ predicts a heatmap $p^{\text{out}}_{x, \tilde{r}} \in \mathbb{R}^{N_{r}\times N_{\varphi}}$, where for each angle $\varphi_i$ a corresponding lumen radius is extracted via a differentiable DSNT module~\cite{dsnt,dsnt-landmarks-gajowczyk}.
These lumen radii are transformed back to Cartesian space to form the predicted lumen contour, hence $h(\cdot; \theta_h)$ can be optimized using a set of ground-truth contours. Network $h(\cdot ; \theta_h)$ is translation equivariant and operates on a polar transformed image, implying rotational equivariance in Cartesian space: a rotation of the 2D input plane leads to the same rotation of the output contour. For the loss function we adopt the same setup as in~\cite{dsnt-landmarks-gajowczyk,dsnt} consisting of a Euclidean distance term between the predicted and ground truth lumen contours and Jensen-Shannon divergence between the $p^{\text{out}}_{x, \tilde{r}}$ and $\text{GT}$ heatmap.

\subsection{Surface reconstruction}
Iterative tracking results in a set of contours describing the lumen surface of vessel $v_i$, but this is not watertight. To reconstruct a watertight surface subject to these contours, we use a neural field (right Fig.~\ref{fig:pipeline}). This neural field consists of a multi-layer perceptron that is conditioned on spatial coordinates $x \in \mathbb{R}^3$ and is optimized to represent the signed distance function of the lumen surface~\cite{ma2021neural}. As these neural fields are conditioned on continuous coordinates, they have an infinite resolution. Moreover, a smooth transition between connecting vessels is made by taking a smoothed minimum of their signed distance functions~\cite{alblas2022going}. From the neural field, we can obtain a mesh model of any resolution using the marching cubes algorithm. 

\subsection{Data}
We use an in-house dataset consisting of 80 pre-operative contrast-enhanced CT (CTA) scans of AAA patients obtained from Amsterdam UMC, containing at least the thoracic region until the iliac bifurcation. In-plane resolutions and slice thicknesses range between 0.6$\times$0.6 and 1.0$\times$1.0 mm\textsuperscript{2} and 0.5-2.0 mm, respectively. For each patient, we obtained manual annotations consisting of centerlines and locally orthogonal contours of the lumen of the abdominal aorta until the top of the T12 vertebra, the common iliac arteries, and the renal arteries.

\section{Experiments \& Results}
We split the dataset into 14 test cases and 66 train cases and use the same data split for both networks in the local vessel segmentation module. We denote the orientation estimation module as $\mathcal{T} = \{ g(\cdot;\theta_g) \}$ and the contour regression module as $\mathcal{S} = \{g(\cdot;\theta_g), h(\cdot;\theta_h)\}$. Note that the orientation estimation and contour regression are trained separately. 
Both $\mathcal{T}$ and $\mathcal{S}$ were trained with an Adam optimizer, with learning rates of $0.005$, $0.001$ and batch sizes of $20$ and $10$ respectively. For both training and inference, we used scales $R=\{ 5, 10, 15, ..., 80 \}$ for the backbone. For tracking we use step sizes $\Delta_{aorta} = 1$ mm, $\Delta_{iliac} = 0.5$ mm, $\Delta_{renal} = 0.3$ mm and delineate contour every $\eta = 5$ steps.

The global controller provided seed points and boundary conditions for tracking the abdominal aorta, iliac arteries and renal arteries. Unless stated otherwise, we used the following heuristics for $p_i$ and $\Omega_i$. $p_{\text{aorta}}$ and $p_{\text{iliac}}$ are the centers of mass from the aorta and iliac masks provided by TotalSegmentator, respectively. $p_{\text{renal}}$ is a bit more challenging to obtain, as TotalSegmentator does not include renal arteries. Instead, we estimate the renal artery by computing the shortest path between the kidneys and the aorta based on image intensities and let $p_{\text{renal}}$ be the center of this path. 
$\Omega_{\text{aorta}}$ consists of the region above the T12 vertebra and the iliac artery segmentations, $\Omega_{\text{iliac}}$ contains the aorta and the region below the L5 vertebra, and $\Omega_{\text{renal}}$ contains the kidneys and the aorta.

\begin{figure}[!t]
    \centering
    \includegraphics[width=\textwidth]{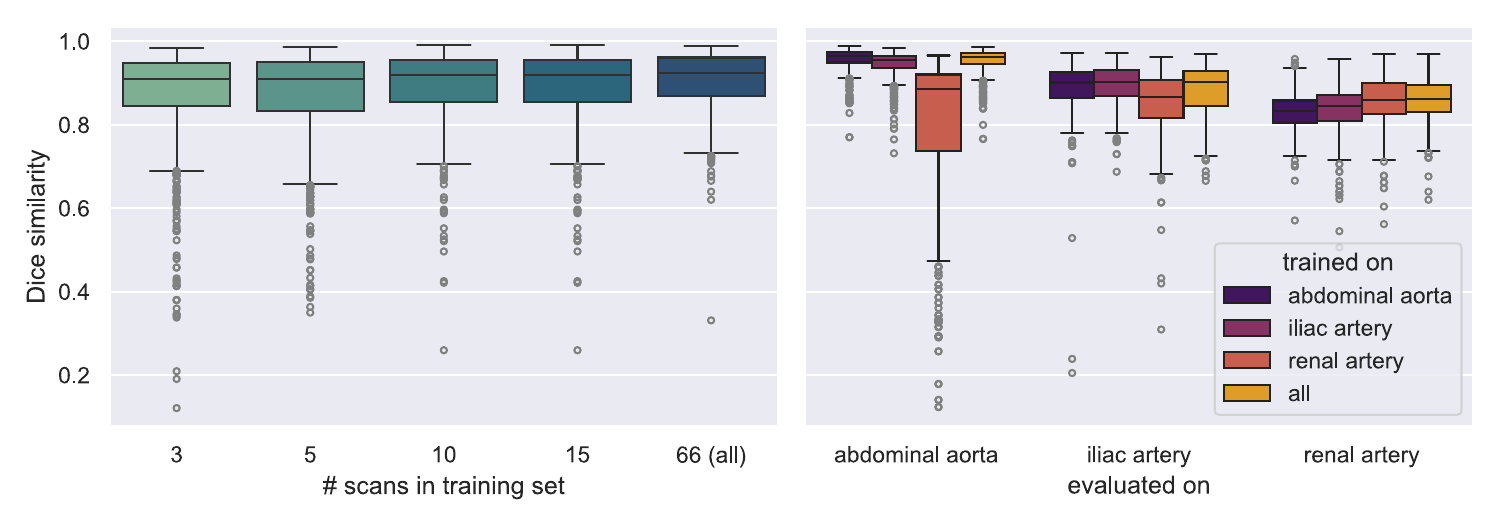}
    \caption{DSC scores of the contour regression modules on 2D contours in the test set for the data ablation experiments. \textit{Left:} DSC for all contours for data ablation at patient level. \textit{Right:} DSC per vessel for data ablation at the vessel level.}
    \label{fig:benchmark}
\end{figure}

\subsection{Segmentation quality}
We evaluate the quality of the automatically extracted 3D vascular models on the test set. 
For each of the 14 test cases, we automatically extract the lumen of the abdominal aorta, renal arteries, and iliac arteries. We compare the quality of these segmentations to an nnU-Net benchmark~\cite{nnunet}. 
We obtained Dice similarity coefficients (DSCs) of $0.84$ $\pm$ $0.05$ and $0.83$ $\pm$ $0.14$ for our method and nnU-Net, respectively, demonstrating that we can fully automatically obtain watertight meshes with sub-voxel accuracy without compromising on overlap metrics. 

\subsection{Data efficiency}
We train our iterative vessel segmentation method using contour annotations on 2D planes in a 3D images. Therefore, a single 3D scan contains hundreds of training samples. This - in combination with the symmetries that we embed - makes our method very data efficient. We retrained the contour regression model on subsets of 3, 5, 10, and 15 patients randomly selected from the training set to assess this data efficiency. We evaluate the performance of each model using the DSC on the 2D contours in the test set. Figure \ref{fig:benchmark} (left) shows our method achieves a median DSC of at least 0.9, even when only three patients were used for training.

As the contour regression module is scale-invariant, it should generalise to vessels of unseen calibre. To test this property, we perform data ablation at the artery level. We retrain three contour regression modules separately on the lumen contours of the abdominal aorta, iliac arteries, and renal arteries, whose diameters range between 20-55 mm, 10-15 mm, and 5-7 mm, respectively. We assess the performance of the four models using the DSC on the 2D contours of the test set. Figure \ref{fig:benchmark} (right) shows that overall, our model can generalise to arteries of unseen size. For example,  the model trained on abdominal aorta contours performs on par with the model trained on iliac arteries when segmenting iliac arteries and vice versa. Moreover, the model trained on all arteries shows performance comparable to the vessel-specific model, indicating that heterogeneity of vessel size in the training data does not deteriorate segmentation performance. However, we observe a performance drop for segmentation of the abdominal aorta for the model trained on renal arteries. 


\begin{figure}[t!]
    \centering
    \includegraphics[width=\textwidth]{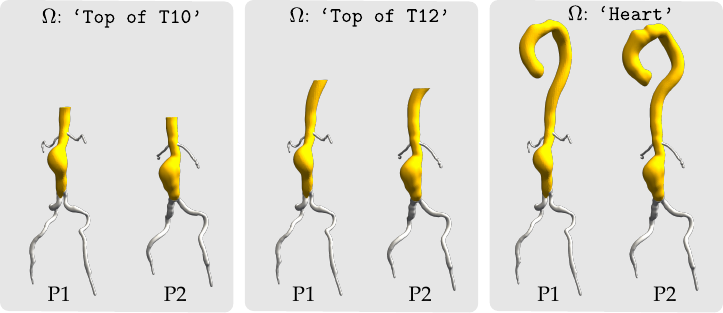}
    \caption{Automatically acquired 3D vascular models using different $\Omega_{\text{aorta}}$.}
    \label{fig:flexible_bcs}
\end{figure}

\subsection{Modifying boundary conditions}
The global controller we present provides flexible inclusion of different vessel ROIs in the 3D model. To show this, we extract multiple 3D models while adapting $\Omega_{\text{aorta}}$. We include three different regions in $\Omega_{aorta}$ to terminate iterative segmentation in the superior direction: until the T10 vertebra, the T12 vertebra, and the heart. Figure \ref{fig:flexible_bcs} shows visualizations of the resulting 3D vascular models. Although the iterative tracker was trained only on contours of the aorta between the T12 vertebra and iliac bifurcation, it still accurately tracks and segments the aortic arch due to its symmetry-preserving properties.

\section{Conclusion \& Discussion}
We have presented a general framework for automatically extracting smooth, topologically correct vascular models, combining strengths of global and local vessel segmentation methods. A global controller provides boundary conditions for a local iterative vessel segmentation model, that exploits local scale- and rotational symmetries to generalise to vessels of unseen sizes and orientations. Our post-processing step provides a neural field of the vascular model that can be reconstructed to a watertight mesh of any desired resolution. We demonstrated the effectiveness of symmetry preservations by automatic extraction of vascular models of abdominal aortic aneurysms, iliac arteries and renal arteries.

By leveraging local scale- and rotational symmetries, our iterative joint tracking and segmentation module generalizes to vessels of sizes (Fig. \ref{fig:benchmark}, left) and segments (Fig. \ref{fig:flexible_bcs}) unseen during training. Despite this strong generalisation, some symmetries cannot be preserved. For example, the contour regression model trained on renal arteries struggles to segment the abdominal aorta lumen if surrounded by a thrombus, as this was not seen during training. 
Moreover, generalisation to other image modalities, e.g. MRI or ultrasound requires retraining. 
However, as the local vessel segmentation step can be trained on contour annotations of a few patients, data collection and annotation costs are low. 

The global controller we present in this work gives control over vessel ROIs to incorporate in the final 3D model. In combination with the generalising properties of the local iterative tracker, this allows us to automatically build 3D vascular models in other anatomical regions by providing a different set of heuristics. We used coarse segmentation masks from TotalSegmentator \cite{total-segmentator} as input for our global controller, an off-the-shelf tool. In combination with the generalising local segmentation method, it has the potential to automatically segment any vascular region without the need for additional specific annotations. A potential drawback of this approach is that the local segmentation model can only segment those vessel segments for which at least a seed point can be identified. We found, that in some cases, the global controller could not identify renal arteries, thereby affecting overall segmentation performance. Note that we use TotalSegmentator but are not dependent on this specific model. Any existing model that provides binary segmentation masks or seed points could be used, and multiple models could be combined for specialized tasks. 

As the boundary conditions for each artery are standardized, the ROIs of the resulting 3D vascular models are consistent across subjects. This enables large-scale studies of the role of hemodynamic and morphological features in the prognosis of cardiovascular diseases and could in future work be extended to quantification of other vessel-specific structures such as intraluminal thrombus and plaques that are important in the analysis of cardiovascular diseases.

In conclusion, we have shown that global constraints can be imposed on local tracking and segmentation models to provide high-quality watertight segmentations of the vasculature.

\subsubsection{Acknowledgments:}
This project has received funding from the European Union's Horizon Europe research and innovation programme under grant agreement No 101080947 (VASCUL-AID).
Jelmer M. Wolterink was supported by the NWO domain Applied and Engineering Sciences Veni grant (18192).

\newpage
\bibliographystyle{splncs04}
\bibliography{biblio}

\end{document}